**Title:** Designing antibiotic cycling strategies by determining and understanding local adaptive landscapes

**Authors:** Christiane P. Goulart[1], Mentar Mahmudi[2], Kristina A. Crona[1], Stephen D. Jacobs[1], Marcelo Kallmann[2], Barry G. Hall[3], Devin C. Greene[1], Miriam Barlow[1]

**Author Affiliations:** [1]School of Natural Sciences, University of California, Merced, , Merced, California, USA

[2]School of Engineering, University of California, Merced, , Merced, California, USA

[3]Bellingham Research Institute, Bellingham, Washington, USA

**Corresponding Author:** Miriam Barlow

University of California, Merced

5200 N. Lake Rd,

Merced, CA. 95343

**Telephone:** 209.228.4174

**Email:** miriam.barlow@gmail.com

**Abstract:** The evolution of antibiotic resistance among bacteria threatens our continued ability to treat infectious diseases.  The need for sustainable strategies to cure bacterial infections has never been greater.  So far, all attempts to restore susceptibility after resistance has arisen have been unsuccessful, including restrictions on prescribing [1] and antibiotic cycling [2,3]. Part of the problem may be that those efforts have implemented different classes of unrelated antibiotics, and relied on removal of resistance by random loss of resistance genes from bacterial populations (drift). Here, we show that alternating structurally similar antibiotics can restore susceptibility to antibiotics after resistance has evolved.  We found that the resistance phenotypes conferred by variant alleles of the resistance gene encoding the TEM β-lactamase ($bla_{TEM}$) varied greatly among 15 different β-lactam antibiotics. We captured those differences by characterizing complete adaptive landscapes for the resistance alleles $bla_{TEM-50}$ and $bla_{TEM-85}$, each of which differs from its ancestor $bla_{TEM-1}$ by four mutations. We identified pathways through those landscapes where selection for increased resistance moved in a repeating cycle among a limited set of alleles as antibiotics were alternated. Our results showed that susceptibility to antibiotics can be sustainably renewed by cycling structurally similar antibiotics. We anticipate that these results may provide a conceptual framework for managing antibiotic resistance.  This approach may also guide sustainable cycling of the drugs used to treat malaria and HIV.

**Introduction:**

For the past seventy years, the world has been flooded with β-lactam antibiotics [4,5]. They have been the favored treatment for most bacterial infections because of their efficiency, specificity, and low toxicity [6,7]. In the 1940s and beyond, penicillin and penicillin derivatives were the most heavily used β-lactams [8]. However, specificity of penicillins for gram positive bacteria and increasing frequencies of β-lactamases in resistant bacteria spurred the development of extended spectrum β-lactams including cephalosporins, monobactams, and carbapenems in the 1980s [5]. Within a few years, resistance to those antibiotics also evolved and the frequencies of those resistance determinants have continued to rise [5,9]. Decreasing the consumption of β-lactams has not been successful in lowering resistance rates [1], nor has alternating (cycling) their use with unrelated (non β-lactam) classes of antibiotics [2,3]. However these attempts to control antibiotic resistance have included *ad hoc* selections of antibiotics, usually with no underlying theoretical or experimental framework.

It is unfortunate that the development of the necessary theoretical and experimental underpinnings of successful antibiotic cycling lagged behind the efforts of the medical community. However, theoretical and experimental work directed at this problem is starting to catch up. Recommendations about how to derive the optimal orders of antibiotics and the duration over which they should be applied have been introduced and are being refined [3,10,11,12]. It is fairly clear at this point that although clinical cycling may not be reliable yet, more informed and sophisticated models have the potential to make management of resistance by antibiotic cycling a robust approach to the resistance problem.

We asked whether alternating the use of structurally similar antibiotics (all β-lactams) might restore their usefulness. We reasoned that when the selective pressure resulting from consumption of an antibiotic is removed from a population, either through cycling or decreased consumption, pleiotropic fitness costs associated with expression of the resistance mechanism will be the major selective pressure removing resistance determinants from bacterial populations. If those fitness costs are extremely low, or if compensatory mutations have ameliorated their effects, such that there are essentially no fitness costs associated with expression of the resistance mechanism, then drift may be the major mechanism for removing those resistance determinants [13,14,15,16,17]. The enormity of bacterial populations and the impossibility of complete discontinuance of an antibiotic make removal of resistance by drift too slow a process to have any practical outcome. Instead, we reasoned that if the selective pressure for the evolution of a specific resistance determinant could be in constant flux, then evolution would occur much more rapidly, and always have a moving target. We wondered whether it might be possible to direct the evolution of resistance in a cyclical fashion.

The experimental model we used to test this approach was the TEM family of β-lactamases. They are often the most frequently encountered resistance genes in clinical bacterial populations. Collectively they confer resistance to the majority of β-lactam antibiotics [9]. Over 200 unique variants of TEM that differ in amino acid sequence have evolved since the gene encoding the TEM-1 β-lactamase (*bla*$_{TEM-1}$) was first identified in 1963 (http://www.lahey.org/Studies/). The consumption of the antibiotics responsible for selecting those substitutions has been recorded
[18,19,20,21,22,23,24,25,26,27,28,29,30,31,32].

**Results**

In this study, we have determined the topologies of adaptive landscapes[33,34,35,36,37,38,39,40,41,42,43,44,45,46,47] that were traversed as two $bla_{TEM}$ alleles evolved naturally. The genes $bla_{TEM-50}$ [48] and $bla_{TEM-85}$ [49] differ from their ancestor $bla_{TEM-1}$ by four mutations that result in amino acid substitutions. Those mutations have arisen independently multiple times during the course of $bla_{TEM}$ evolution [50] and confer adaptive benefits. Although those mutations have adaptive roles in certain genetic backgrounds and selective environments, they are not always beneficial in every genetic background. This phenomenon is called sign epistasis. To characterize those landscapes, we created all possible combinations of the mutations found in $bla_{TEM-50}$ and $bla_{TEM-85}$ (Table 1), and determined the resulting resistance phenotypes by disk diffusion testing with 15 β-lactam antibiotics (Table 2) that have been used heavily during the period in which TEM variants have arisen.

We assumed the strong selection weak mutation (SSWM) model [51] as both alleles evolved; we also assumed that increased resistance indicates increased fitness [52]. We organized our results using fitness graphs (SI, Figures S1-S30) where 16 nodes correspond to the 16 alleles. In principle, a fitness graph coincides with the Hasse-diagram of the power set of events wherein addition fitness differences between adjacent genotypes, ie genotypes that differ by one mutation, are indicated (see materials and methods). Specifically, the wild-type allele ($bla_{TEM-1}$) is at the bottom, alleles with single mutations are one level above, alleles with two mutations are at the next level, followed by those with three and then four mutations (ie $bla_{TEM-50}$ and $bla_{TEM-85}$). Line segments drawn between adjacent nodes complete the graph. Green lines indicate selection for mutation, red lines indicate selection for reversion; absence of a line indicates that the adjacent nodes are phenotypically equivalent. We used one-way ANOVA testing to determine 95% confidence intervals around the mean resistance phenotypes and to assign direction. The fitness graphs reflect coarse properties of the fitness landscapes, including accessible trajectories, the number of peaks and sign epistasis. Such properties are important for our approach to drug cycling programs. However, since fitness graphs depend on fitness ranks of genotypes only, they will not reflect quantitative aspects which may be relevant for recombination. Fitness graphs reveal the adaptive potential under the assumption that no recombination, double mutations or other extreme genetic events take place.

**The Complexity of Fitness Landscapes**

Additive fitness landscapes have a single peak. In contrast, random (uncorrelated or rugged) fitness landscapes have no correlation between the fitness of adjacent alleles. Random fitness and additivity can be considered as two extremes with regard to the amount of structure in the fitness landscapes. The degree of additivity, roughly how close the landscape is to an additive landscape, is of interest for comparing landscapes in different settings. Since we work with qualitative information we use the qualitative measure of additivity, a value ranging from 0 to 1 for fitness landscapes, where the value is 1 for additive landscapes and close to 0 for a random fitness landscape [53]. The mean values were 0.33 for TEM-50 and 0.57 for TEM-85, using the landscapes where the measure applies. The complete statistics show that the TEM-85 landscapes have considerably less additivity than for additive landscapes and considerably more additivity than expected for random fitness landscapes. The results for TEM-50 point in the same direction (see materials and methods).

In an additive landscape where TEM-85 confers the greatest fitness advantage, fitness should always increase with the addition of more mutations. However, in the cefotaxime and ceftazidime landscapes where TEM-85 does confer the greatest fitness advantage, there are seven instances in each where fitness increases via reversion of a previously existing mutation. Overall, there are several occurrences of sign epistasis in 14 out of 15 landscapes. The cefoxitin landscape (Figure S26) has nearly no sign epistasis because it is almost flat, in the sense that there are no significant fitness differences for most alleles. For TEM-50 sign epistasis also occurred in 14 out of 15 landscapes and the ampicillin landscape was flat (Figure S1).

**Adaptive trajectories in single-antibiotic and fluctuating environments**

*TEM-50 landscapes*

In the 15 adaptive landscapes that include $bla_{TEM-50}$ related alleles, there were no pathways containing consecutive increases of resistance between $bla_{TEM-1}$ and $bla_{TEM-50}$ (See figures S1-S15). Based on this result, it is possible that recombination occurred during the evolution of $bla_{TEM-50}$. However, an alternate explanation for the evolution of $bla_{TEM-50}$ is fluctuation of environments as different antibiotic have been administered. When the results from the 15 different landscapes were simultaneously considered, we identified 5589 trajectories between $bla_{TEM-1}$ and $bla_{TEM-50}$.

*TEM-85 landscapes*

In contrast, two of the 15 adaptive landscapes that include $bla_{TEM-85}$ related alleles contain pathways of consecutively increasing resistance between $bla_{TEM-1}$ and $bla_{TEM-85}$ (Figures S16-S30). Cefotaxime (Figure 1a) and ceftazidime (Figure 1b) can individually select for the evolution of $bla_{TEM-85}$ with either two or three pathways, respectively. TEM-85 is the allele of greatest fitness for both cefotaxime and ceftazidime. The cefotaxime landscape has 2 peaks and the ceftazidime landscape has 4 peaks. We computed the probabilities for a population going to fixation at TEM-85, rather than at suboptimal peaks, using a basic model which assumes that available beneficial mutations are equally likely to occur and go to fixation. For cefotaxime the probability for fixation at TEM 85 was 75%, and for ceftazidime 12.5%. When all landscapes were simultaneously considered, which is appropriate under circumstances of fluctuating selection, we found 15,716 pathways between $bla_{TEM-1}$ and $bla_{TEM-85}$.

These results are consistent with Weinreich *et al.* [52] in that when a single environment is considered, there are few pathways through which evolution can proceed. These results are also consistent with the study by Bergstrom *et al.* [3] in which they found that random fluctuations of antibiotics can accelerate the evolution of resistance.

**Antibiotic Cycles**

The complexity of the adaptive landscapes makes it possible to identify cycles of antibiotics that are likely to effectively manage resistance. The somewhat frequent increases in resistance that result from reversions of mutations indicate that the evolution of resistance is sometimes reversible. Based on this observation we investigated approaches for cycling antibiotics as a method for managing resistance. We determined whether it was possible to cyclically restore susceptibility to a sequence of antibiotics by alternating exposure to those antibiotics as follows: We first identified the alleles that were local optima in the presence of at least one antibiotic, and that were also adaptive valleys in the presence of at least one other antibiotic. Next we identified the antibiotic environments where those local optima and valleys exist. We then identified pathways that formed closed loops within those adaptive landscapes; the pathways returned to the allele where they had begun. For $bla_{TEM-50}$ landscapes the antibiotics we identified were cefepime, cefprozil, and ceftazidime (Figure 2). We found that 41,961 different cycles exist in those landscapes. For the $bla_{TEM-85}$ landscapes, we identified the antibiotics cefprozil, ceftazidime, cefotaxime, and ampicillin as the most appropriate choices for antibiotic cycling and with those, we identified 1770 cycles (Figure S31). These results indicate that there are numerous routes for resistance to be reversed when those three antibiotics are cycled, which is an indication that this approach is robust. If the order of antibiotics is perturbed, the effects of cycling those antibiotics should be consistent. One caveat is that in the case of TEM-85, there are very few reversions that increase fitness when the allele $bla_{TEM-85}$ has been reached (Figure S31), allowing the potential for "escape" from the cycling regimen. Generating adaptive landscapes for more antibiotics may ameliorate this situation.

**Discussion**

Our results indicate that the occurrence of sign epistasis may provide a means for sustainably renewing the usefulness of antibiotics once resistance to them has evolved. Historic failures of *ad hoc* cycling programs for antibiotics in hospitals have no bearing on our approach. The scheme for drug cycling we suggest relies on current laboratory techniques, as well as established theory of adaptation, and it remains to evaluate our approach in a clinical setting. Our results indicate that abundant sign epistasis exists for the TEM resistance determinants and that it provides a means for sustainably renewing the usefulness of β-lactam antibiotics once resistance to them has evolved. An obvious limitation in our approach is that we have considered only a few mutations associated with antibiotic resistance. For practical solutions, a more complete picture is required. Other antibiotic cycling studies have added significantly towards our understanding of what factors will improve cycling. A knowledge of pleiotropic fitness costs associated with resistance mechanisms can help to inhibit the evolution of multi-drug resistant strains and possibly eliminate those that already exist. The order and timing in which antibiotics are applied also have a significant effect on the occurrence of resistance. Additionally, a recent study that demonstrated the effectiveness of a program in which a hospital cycled among β-lactam antibiotics to reduce resistance over a period of several years [54]. This success is consistent with our results and may have benefitted in its design from the apparent absence of pleiotropic fitness costs associated with expression of most serine β-lactamases [55].

### Recommendations for further development of cycling

It is likely that effective cycling programs will be specific to local environments and the specific resistance alleles in that environment. The identification of an effective antibiotic cycling program will require identifying the resistance alleles currently circulating in the local environment, determining the fitness of each of those alleles with respect to the set of antibiotics being considered, then identifying those drugs that will result in a repeating cyclic path through the local adaptive landscape. Additionally, optimizing the order in which antibiotics are applied [11,12] and the duration for which the antibiotics are administered are key [3,10].

The required experiments are neither difficult nor time consuming. The analyses will be facilitated by user-friendly programs that are under development. Although the introduction of unexpected novel alleles, either by mutation or from sources external to the local environment, may disrupt an effective cycling program, identification of those alleles and their addition to the analysis is likely permit effective modification of the cycling program. Applying the suggested antibiotic cycling scheme will be made even more practical as it builds upon thorough analyses of many adaptive landscapes. Moreover, in pathogens such as malaria and HIV where fewer and more predictable resistance mechanisms exist, this method may be more easily implemented.

**Methods**

*Mutants and Susceptibility Testing*

We used QuikChange® site-directed mutagenesis (Stratagene) to generate all mutant constructs from the *bla*<sub>TEM-1</sub> gene in plasmid pBR322 (Table 1). Those mutant alleles were expressed in *Escherichia coli* strain DH5-αE. We performed Kirby Bauer Disk Diffusion Susceptibility testing [56] for 10 replicates of each of the 32 strains under 15 commonly prescribed β-Lactam antibiotics (Table 2).

*Identifying Paths and Cycles*

In order to compute all possible combinations of pathways, we have represented the fitness graph of a drug as a possibly cyclic directed graph $G = (N, E)$. The nodes $N$ of the graph represent the alleles, and the directed edges $E$ of the graph are determined by the statistical analysis of the resistance differences among alleles. Nodes $n_i \in N$ do not have costs associated to them, however costs are associated with each edge $e_i \in E$ and are determined by the resistance difference between the two nodes that are connected by the edge.

Since we consider a biallelic system, the genotypes can be represented by a string of 0's and 1's, where the zero-string 0000 represents the wild-type. A fitness graph compares the fitness ranks of mutational neighbors. Roughly, consider the zero-string as the starting point and each non-zero position of a string as an event, i.e., that a mutation has occurred. Under these assumptions the fitness graph coincides with the Hasse-diagram of the power set of events, except that each edge in the Hasse-diagram is replaced with an arrow toward the string with greater fitness.

For a formal definition, a fitness graph for a biallelic L-loci population is a directed graph where each node corresponds to a string (which represents a genotype). The fitness graphs has $L+1$ levels. Each string such that $\sum s_i = l$ corresponds to a node on level $l$ in the fitness graph. In particular, the node representing the zero-string is at the bottom, the nodes representing strings with exactly one non-zero position are one level above, the nodes representing strings with exactly two non-zero positions are on the next level, and the 1-string is at the top. Moreover, the nodes are ordered from left to right according to the lexicographic order where 1 > 0 of the corresponding strings. A directed edge connects each pair of nodes such that the corresponding strings differ in exactly one position. The edge is directed toward the node representing the more fit of the two genotypes.

Fitness graphs reflect coarse properties of fitness landscapes, including sign epistasis. For a complete analysis of fitness landscapes other methods are necessary. The most fine-scaled approach to epistasis is the geometric theory of gene interaction [44].

Once graph $G = (N, E)$ is built and the edge costs are computed and associated with their corresponding edges, path queries in the graph can then be computed. To compute all the pathways, from the initial starting node (0000 in our trials), a search expansion is performed by adding each connected node as a child to the current node in a search tree representation. Since backward edges are possible, a mechanism to detect cycles is included by making sure that expanded nodes do not appear twice in a pathway. When a cycle is detected, the last node of the cycle is not expanded. The overall expansion continues until all the search tree branches reach the target node 1111. The final search tree will represent all possible pathways.

In addition to the search tree, we use a priority queue Q that maintains ranked all the current leaves of the search tree to be expanded. We use a Dijkstra-like cost-to-come sorting function in Q, which represents the accumulated costs of the pathways since the source node. The priority queue ranks all available leaf nodes to be expanded and the node with lowest cost is always expanded first. This guarantees that each node is reached in the order of appearance in the shortest path from the source node to it. This guarantees that the shortest cycles are always found first. Since cycle determination is important in our research, all identified cycles are stored and saved for later analysis.

Although our experiments involved searches with different graphs (single or multiple drugs), searches with either forward or backward edges and searches with different starting nodes (1111 for backwards pathways), the described search method was the same and handled well all situations. The used notation in our figures shows backward edges in red and forward edges in green.

*Degree of additivity:*

The degree of additivity, roughly how close a landscape is to being a completely additive landscape, can be measured in different ways. We used the qualitative measure of additivity which ranges from 0 to 1 for fitness landscapes. For a formal definition: The set $B_p$ consist of all double mutants such that both corresponding single mutations are beneficial.

The set $B \subseteq B_p$ consists of all double mutants in $B_p$ which are more fit than at least one of the corresponding single mutants.

The qualitative measure of additivity is the ratio $\frac{|B|}{|B_p|}$.

$\frac{|B|}{|B_p|} = 1$ for an additive landscapes. For random fitness landscapes, the measure is expected to be close to zero in this setting. Indeed, using standard arguments in the Orr-Gillespie approach, the wild-type has very high fitness also in the new environment, in comparison with a randomly generated genotype. By definition, fitness is uncorrelated for a random fitness landscape, so that double mutants combining beneficial mutations are expected to be no more fit than randomly generated genotypes. It follows that the qualitative measure is close to zero, and a more precise estimate is that it should be less than 3% for random landscapes in this context. The derivation of this result will be published elsewhere. The conclusion depends on an analysis of TEM data from the record of clinically found mutants.

For the 15 TEM-85 landscapes, the qualitative measure applies for 9 out of the 15 landscapes. The result is 0, 0, 0, 1/3, 5/6, 1, 1, 1, 1. The mean value is 0.57. This result deviates considerably from expectations for additive and random landscapes. Indeed, if all landscapes were additive, the result should be 1 in each case modulo measurement errors. For random landscapes, non-zero values are expected to be rare. For TEM-50 the qualitative measure applies for 3 landscapes out of 15 and the corresponding data is 0,0,1. The mean value is 0.33. From the qualitative measure alone, we have an indication that the landscapes are neither all additive, nor all random, also for TEM-50 (even if the data set is small).

The qualitative measure of additivity is useful for comparing a fitness landscape with other empirical landscapes, as well as with additive and random (or uncorrelated) landscapes. The measure is robust in the sense that small differences in the environment, such as (moderate) changes of the concentration of antibiotics, have no impact. Quantitative measures may be more sensitive. However, one should not over interpret the qualitative measure. This is a coarse measure, since it depends on fitness ranks only.

*Probabilities*

By the SSWM assumption we were able to consider fixation of beneficial mutations as independent events. Therefore, we computed the probability for a trajectory as the product of the probabilities of its steps. The probabilities for single substitutions can be determined by the following well-established estimate:

The probability that a beneficial mutation j will be substituted at the next step in adaptation is:

$$\frac{s_j}{s_1 + ... + s_k}$$

where $s_r$ is the fitness contribution of mutation $r$ and where there are $k$ beneficial mutations in total. However, we used the simplified assumption that fitness is equal for available beneficial mutations, so that this probability equals $\frac{1}{k}$.

**Figure Legends**

**Figure 1. Adaptive landscapes of TEM-85.** These diagrams show the pathways through which the $bla_{TEM-85}$ can evolve in a single antibiotic.

1a. (Left) The TEM-85 adaptive landscape in cefotaxime with pathways to TEM-85 indicated.

1b. (Right) The TEM-85 adaptive landscape in ceftazidime with pathways to TEM-85 indicated.

**Figure 2. Example of one possible outcome from antibiotic cycling.**

These diagrams show that by alternating the antibiotics cefepime, ceftazidime, and cefprozil susceptibility to those antibiotics can be restored in bacterial populations expressing variant alleles present in TEM-50 adaptive landscapes.

2a. (Top left) The TEM-50 adaptive landscape in cefepime. Yellow peaks indicate the adjacent alleles that are important during cefepime selection.

2b. (Top right) The TEM-50 adaptive landscape in ceftazidime. Orange peaks indicate the adjacent alleles that are important during ceftazidime selection.

2c. (Bottom left) The TEM-50 adaptive landscape in cefprozil. Red peaks indicate the adjacent alleles that are important during cefprozil selection.

2d. (Bottom right) Composite cycle: The yellow arrow indicates the direction of selection in the presence of cefepime. The red arrows indicate the direction of selection in the presence of cefprozil. The orange arrow indicates the direction of selection in the presence of ceftazidime. Rotation of these antibiotics results in cyclical renewal of antibiotic susceptibility.

**Table 1. Constructs containing all possible combinations of the four mutations found in *bla*<sub>TEM-50</sub> and *bla*<sub>TEM-85</sub>.**

| Number of Substitutions | Binary Allele Code | Variants with mutations found in *bla*<sub>TEM-50</sub> | Variants of mutations found in *bla*<sub>TEM-85</sub> |
|---|---|---|---|
| 0 | 0000 | No Mutations<br>TEM-1 | No Mutations<br>TEM-1 |
| 1 | 1000 | M69L<br>(TEM-33) | L21F<br>(TEM-117) |
| 1 | 0100 | E104K<br>(TEM-17) | R164S<br>(TEM-12) |
| 1 | 0010 | G238S<br>(TEM-19) | E240K<br>(Not identified) |
| 1 | 0001 | N276D<br>(TEM-84) | T265M<br>(Not identified) |
| 2 | 1100 | M69L<br>E104K<br>(Not identified) | L21F<br>R164S<br>(TEM-53) |
| 2 | 1010 | M69L<br>G238S<br>(Not identified) | L21F<br>E240K<br>(Not identified) |
| 2 | 1001 | M69L<br>N276D<br>(TEM-35) | L21F<br>T265M<br>(TEM-110) |
| 2 | 0110 | E104K<br>G238S<br>(TEM-15) | R164S<br>E240K<br>(TEM-10) |
| 2 | 0101 | E104K<br>N276D<br>(Not identified) | R164S<br>T265M<br>(Not identified) |
| 2 | 0011 | G238S<br>N276D<br>(Not identified) | E240K<br>T265M<br>(Not identified) |
| 3 | 1110 | M69L<br>E104K<br>G238S<br>(Not identified) | L21F<br>R164S<br>E240K<br>(TEM-102) |
| 3 | 1101 | M69L<br>E104K<br>N276D<br>(Not Identified) | L21F<br>R164S<br>T265M<br>(Not identified) |
| 3 | 1011 | M69L<br>G238S<br>N276D<br>(Not identified) | L21F<br>E240K<br>T265M<br>(Not identified) |
| 3 | 0111 | E104K<br>G238S<br>N276D<br>(Not identified) | R164S<br>E240K<br>T265M<br>(Not identified) |
| 4 | 1111 | M69L<br>E104K | L21F<br>R164S |

| | | G238S<br>N276D<br>(TEM-50) | E240K<br>T265M<br>(Not identified) |

**Table 2 The antibiotics used to characterize adaptive landscapes.** While not a comprehensive listing of all β-lactam antibiotics, this set contains many heavily used antibiotics and provides good general coverage of β-lactams.

| Antibiotic | FDA approval | Antibiotic Group |
|---|---|---|
| Ampicillin (AM) | 1963 | Penicillin derivative |
| Cefoxin(FOX) | 1978 | Cephalosporin |
| Cefaclor(CEC) | 1979 | Cephalosporin |
| Cefotaxime (CTX) | 1981 | Cephalosporin |
| Ceftizoxime (ZOX) | 1983 | Cephalosporin |
| Cefuroxime (CXM) | 1983 | Cephalosporin |
| Ceftriaxone(CRO) | 1984 | Cephalosporin |
| Amoxicillin +Clavulanic acid (AMC) | 1984 | Penicillin derivative + β-Lactamase inhibitor |
| Ceftazidime (CAZ) | 1985 | Cephalosporin |
| Cefotetan (CTT) | 1985 | Cephalosporin |
| Ampicillin + Sulbactam (SAM) | 1986 | Penicillin derivative + β-Lactamase inhibitor |
| Cefprozil (CPR) | 1991 | Cephalosporin |
| Cefpodoxime (CPD) | 1992 | Cephalosporin |
| Pipercillin + Tazobactam (TZP) | 1993 | Penicillin derivative + β-Lactamase inhibitor |
| Cefepime(FEP) | 1996 | Cephalosporin |

**Supporting Information Legends**

**Figures S1-S15**

**Figures of TEM-50 Adaptive Landscapes**

Legend: Ovals represent alleles. The names are given in binary code (See table 1). The absence of lines indicates no significant difference in resistance phenotypes. Green lines indicate an increase in resistance resulting from addition of a mutation. Red lines indicate an increase in resistance resulting from reversion.

**Figures S16-S30**

**Figures of TEM-85 Adaptive Landscapes**

Legend: Ovals represent alleles. The names are given in binary code (See table 1). The absence of lines indicates no significant difference in resistance phenotypes. Green lines indicate an increase in resistance resulting from addition of a mutation. Red lines indicate an increase in resistance resulting from reversion.

**Figure S31. Example of one possible outcome from antibiotic cycling.**

a. (Top left) Composite cycle: The red arrow indicates an allele (T265M) that will be selected by cefotaxime. The green arrows indicate alleles that will be selected by ceftazidime. The yellow arrow indicates alleles that will be selected by cefprozil.

b. (Top right) The TEM-85 adaptive landscape in cefotaxime. Red peaks indicate the adjacent alleles that are important during cefotaxime selection.

c. (Bottom left) The TEM-85 adaptive landscape in ceftazidime. Green peaks indicate the adjacent alleles that are important during ceftazidime selection.

d. (Bottom right) The TEM-85 adaptive landscape in cefprozil. Yellow peaks indicate the adjacent alleles that are important during cefprozil selection.

**Figure 1**

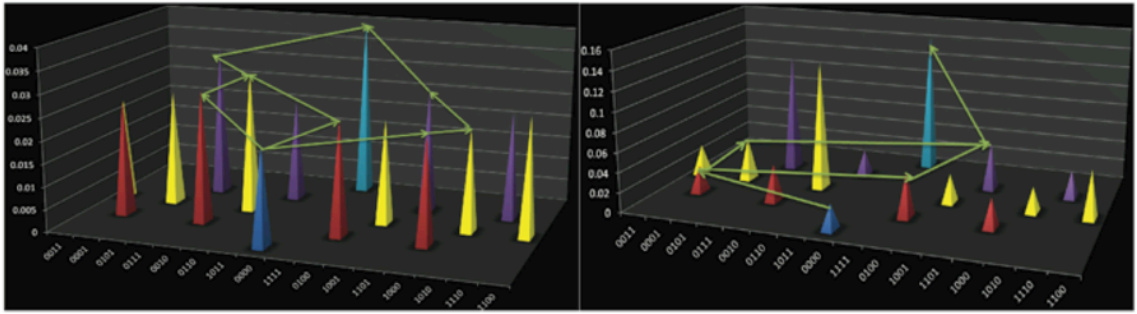

**Figure 2**

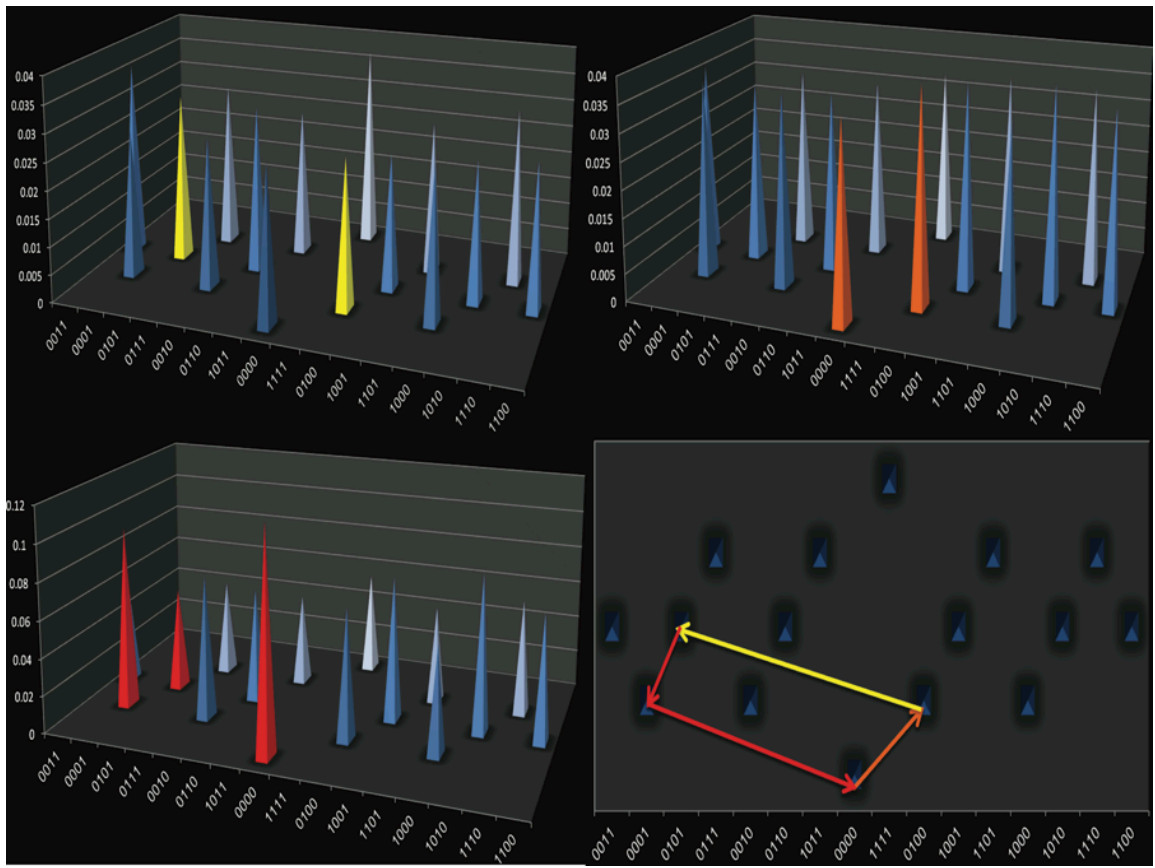

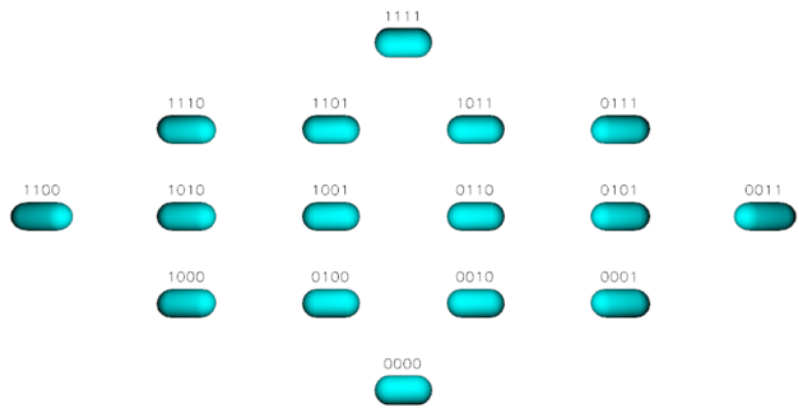

Figure S1 TEM 50 Landscape for Ampicillin

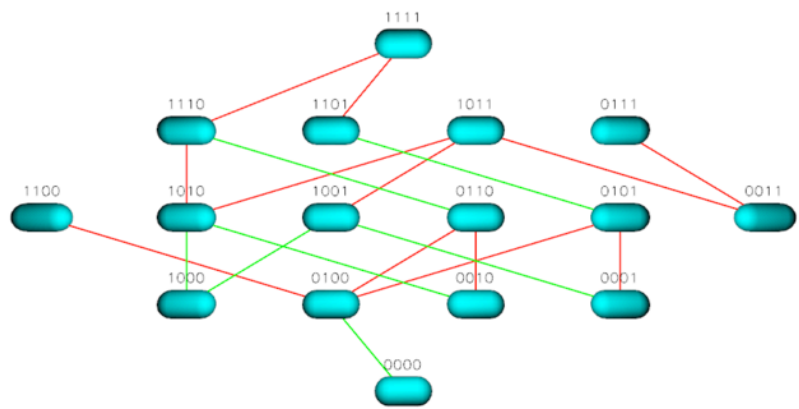

Figure S2 TEM 50 Landscape for Ceftazidime

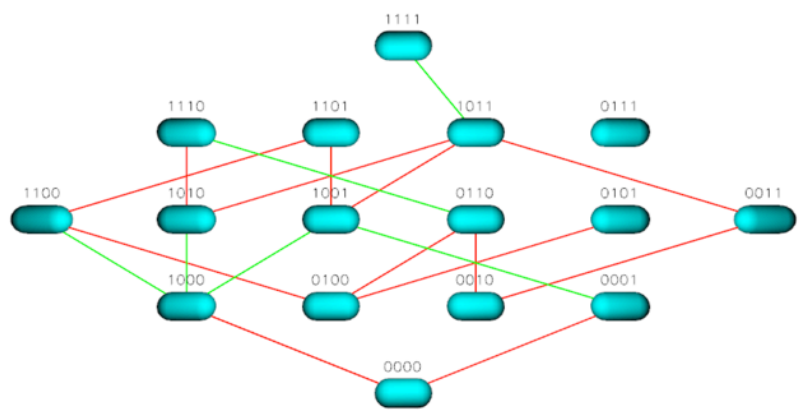

Figure S3 TEM 50 Landscape for Cefaclor

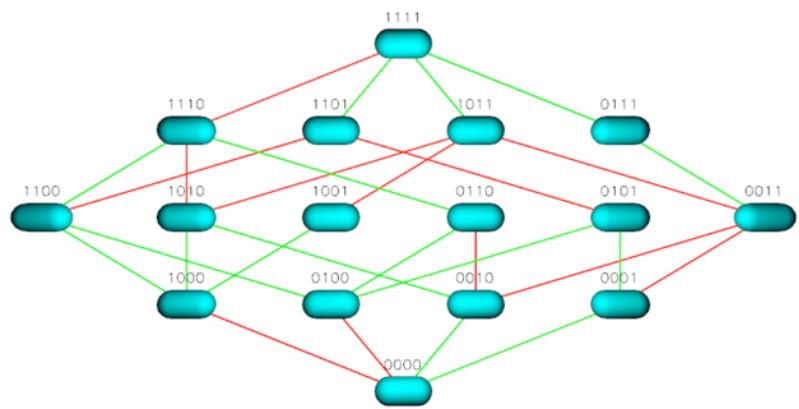

Figure S4 TEM 50 Landscape for Cefpodoxime

Figure S5 TEM 50 Landscape for Ceftriaxone

Figure S6 TEM 50 Landscape for Cefprozil

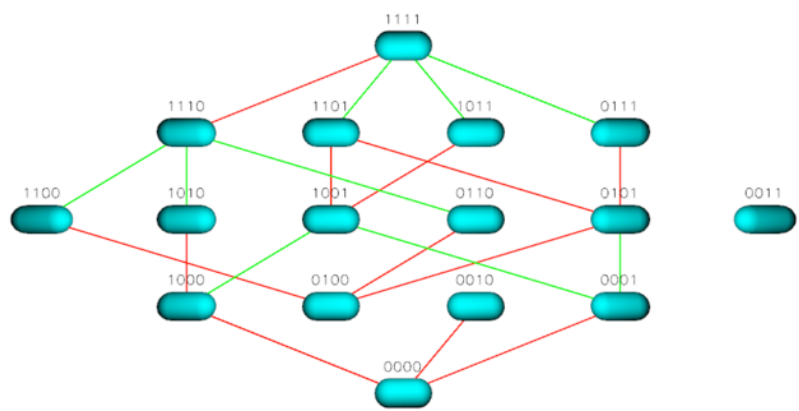

Figure S7 TEM 50 Landscape for Cefotetan

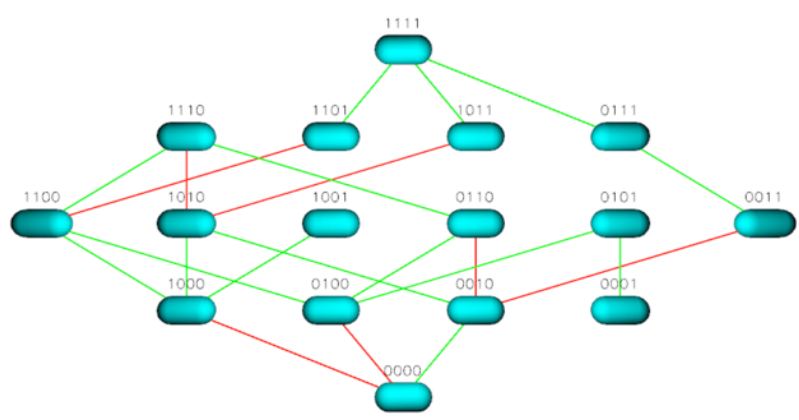

Figure S8 TEM 50 Landscape for Cefotaxime

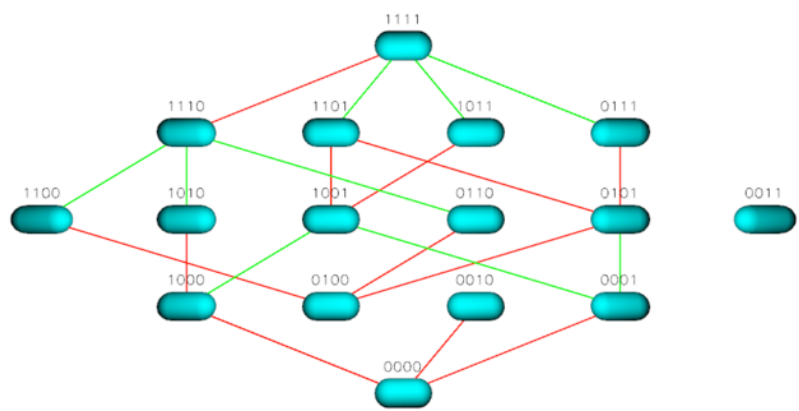

Figure S9 TEM 50 Landscape for Cefuroxime

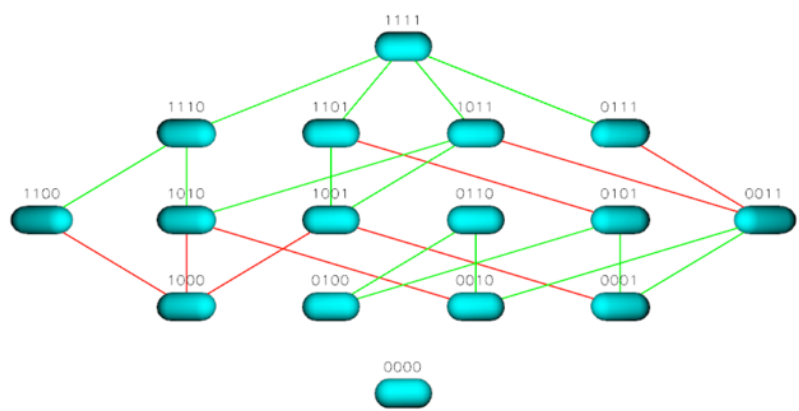

Figure S10 TEM 50 Landscape for Cefepime

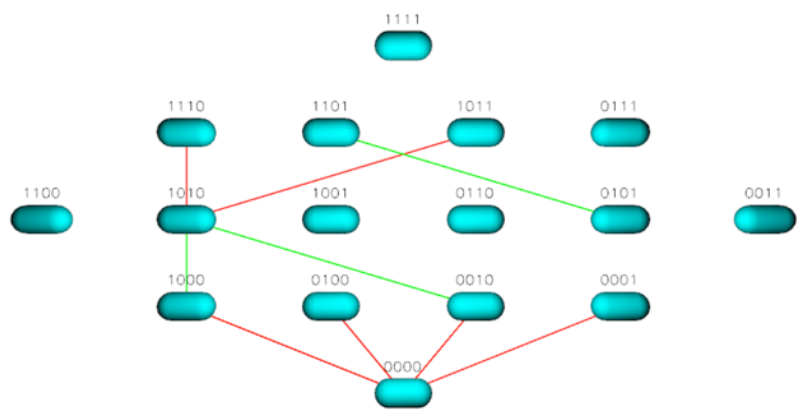

Figure S11 TEM 50 Landscape for Cefoxitin

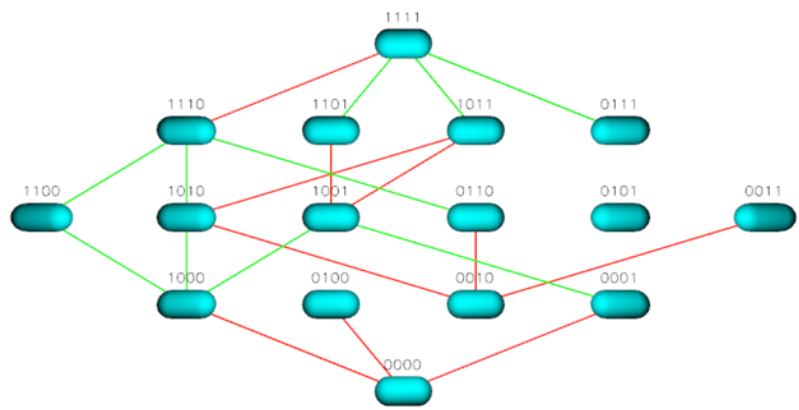

Figure S12 TEM 50 Landscape for Ceftizoxime

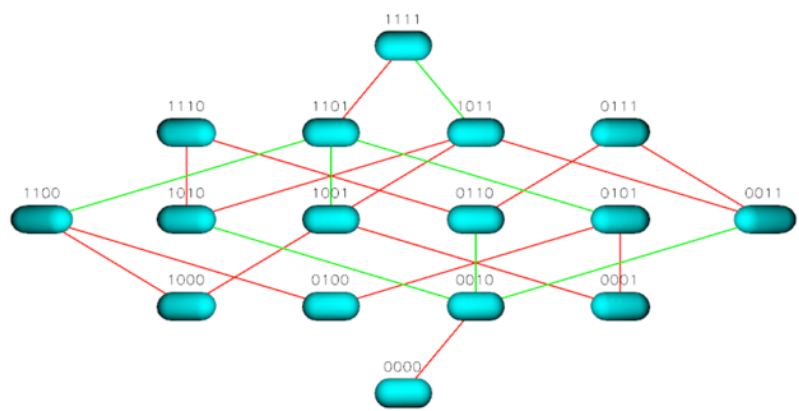

Figure S13 TEM 50 Landscape for Ampicillin + Sulbactam

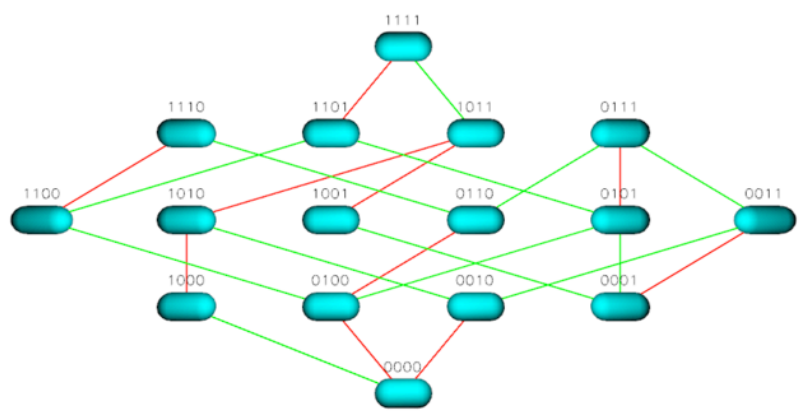

Figure S14 TEM 50 Landscape for Pipercillin + Tazobactam

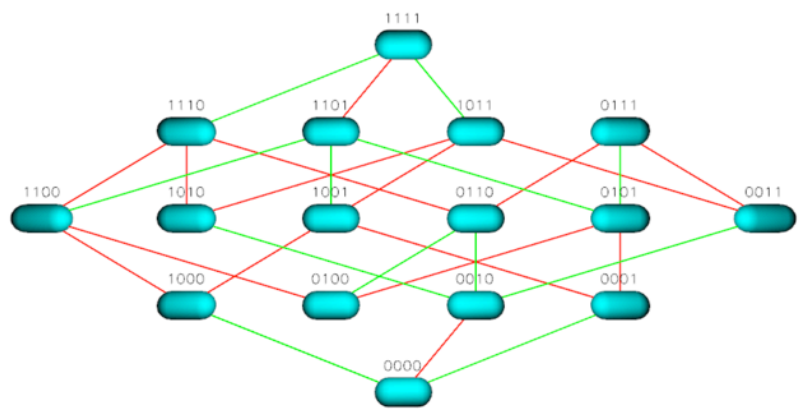

Figure S15 TEM 50 Landscape for Amoxicillin + Clavulanate

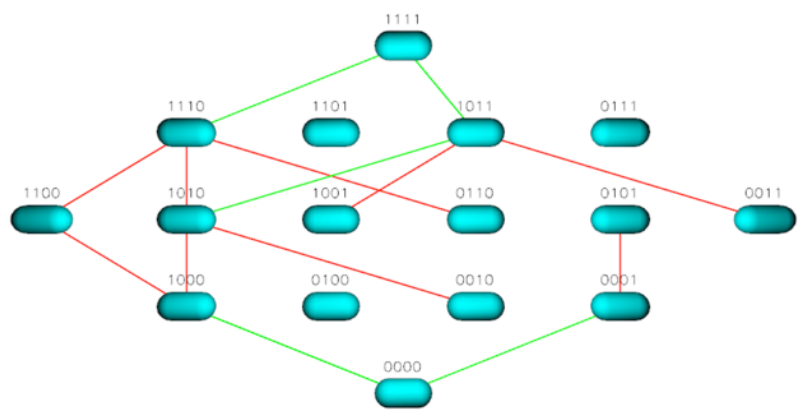

Figure S16 TEM 85 Landscape for Ampicillin

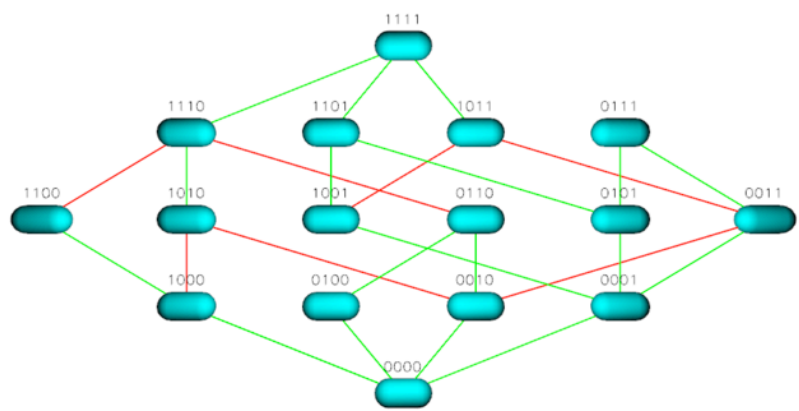

Figure S17 TEM 85 Landscape for Ceftazidime

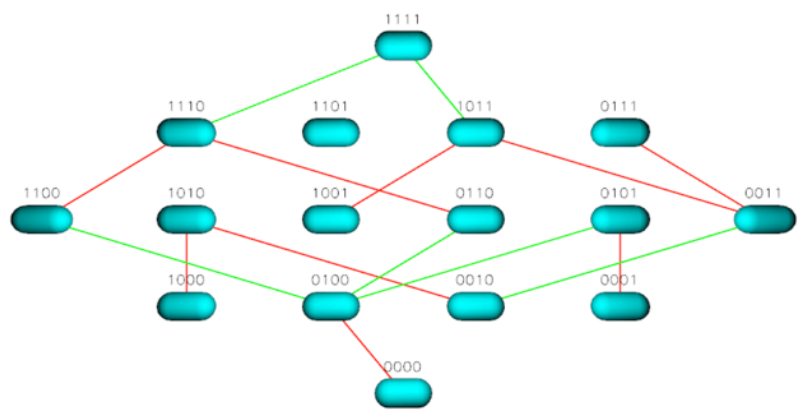

Figure S18 TEM 85 Landscape for Cefaclor

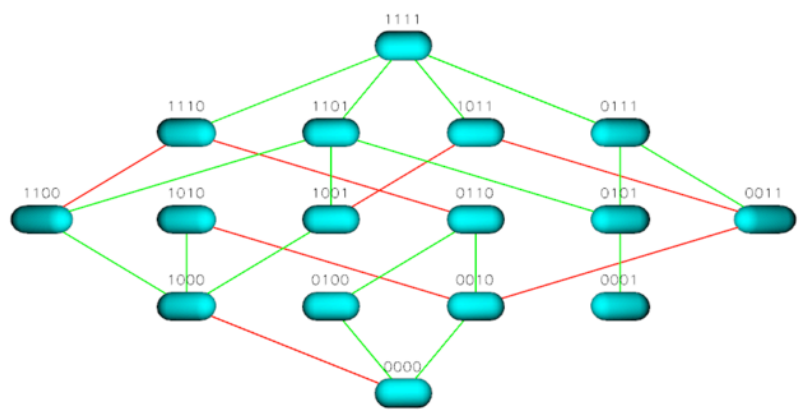

Figure S19 TEM 85 Landscape for Cefpodoxime

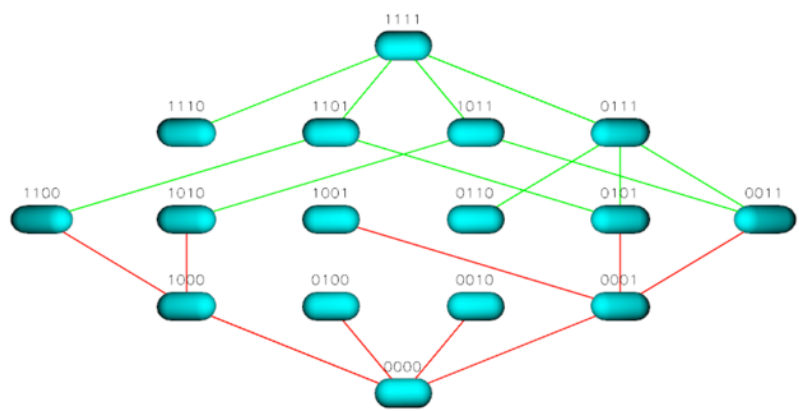

Figure S20 TEM 85 Landscape for Ceftriaxone

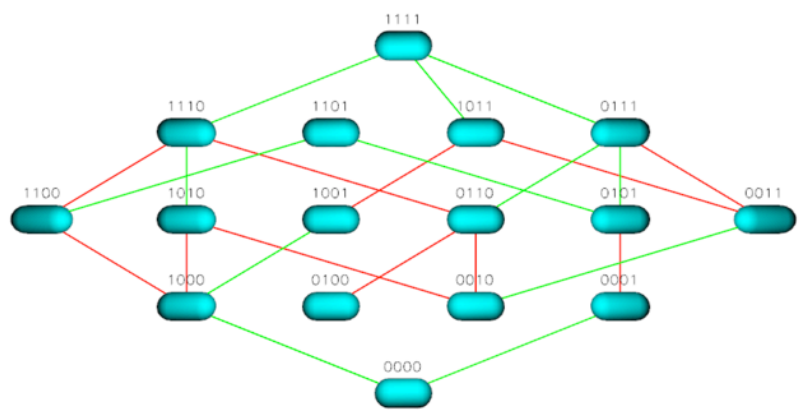

Figure S21 TEM 85 Landscape for Cefprozil

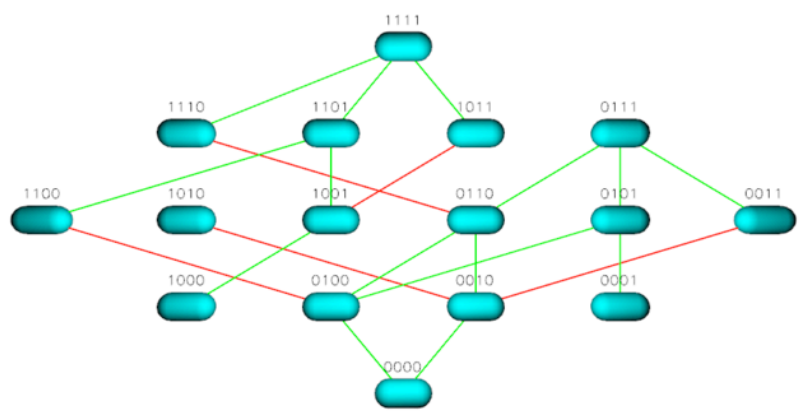

Figure S22 TEM 85 Landscape for Cefotetan

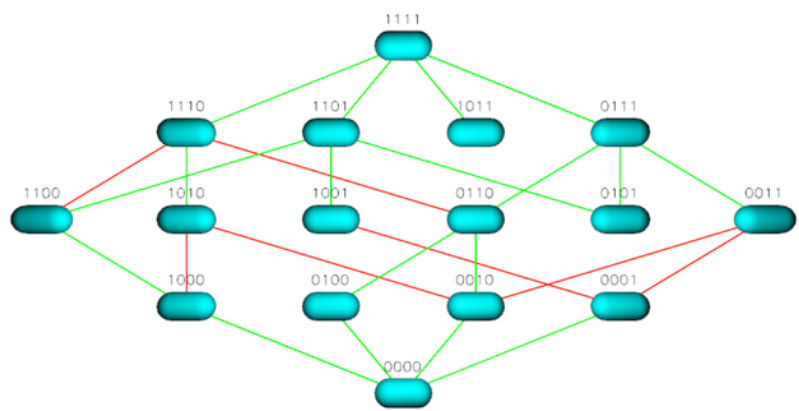

Figure S23 TEM 85 Landscape for Cefotaxime

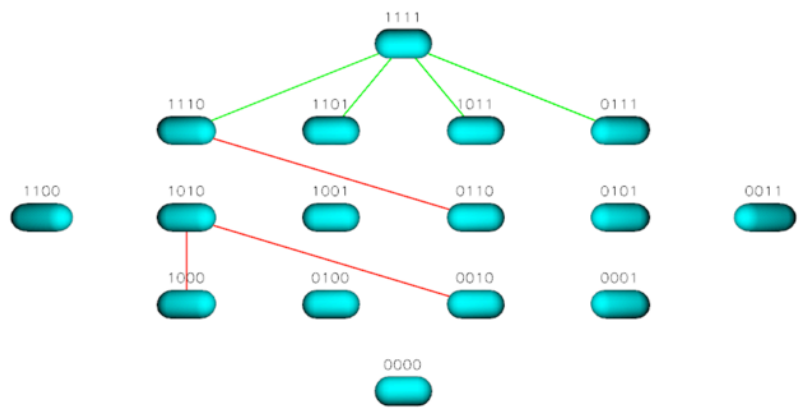

Figure S24 TEM 85 Landscape for Cefuroxime

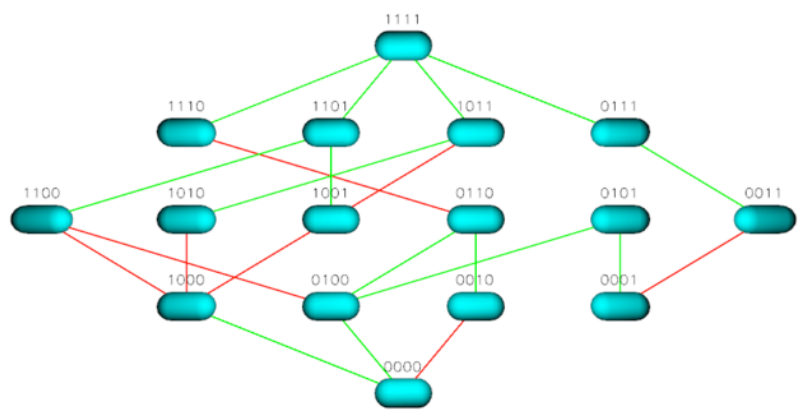

Figure S25 TEM 85 Landscape for Cefepime

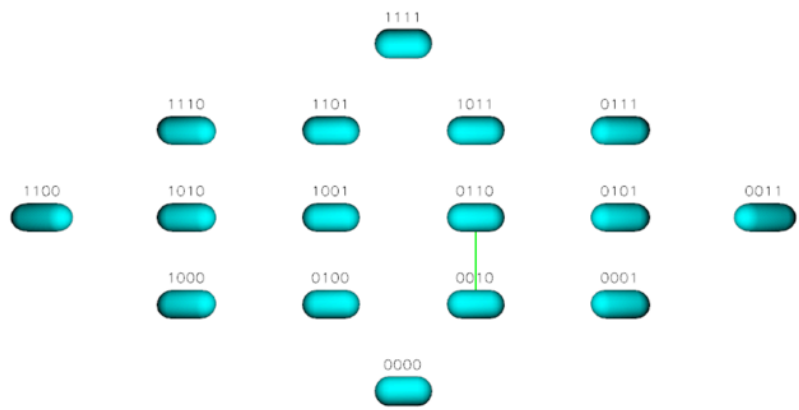

Figure S26 TEM 85 Landscape for Cefoxitin

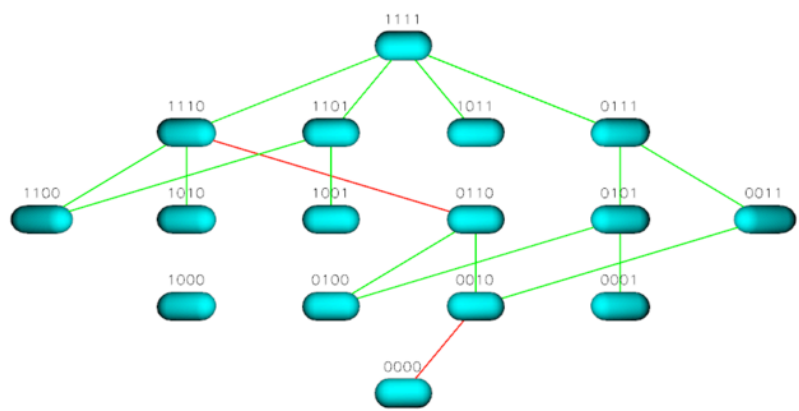

Figure S27 TEM 85 Landscape for Ceftizoxime

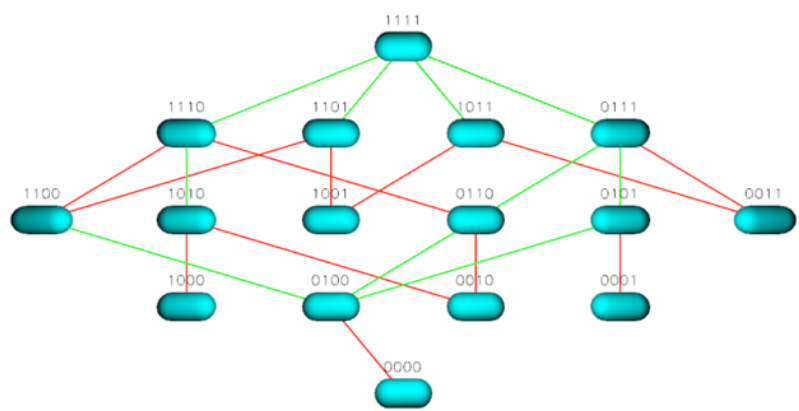

Figure S28 TEM 85 Landscape for Ampicillin + Sulbactam

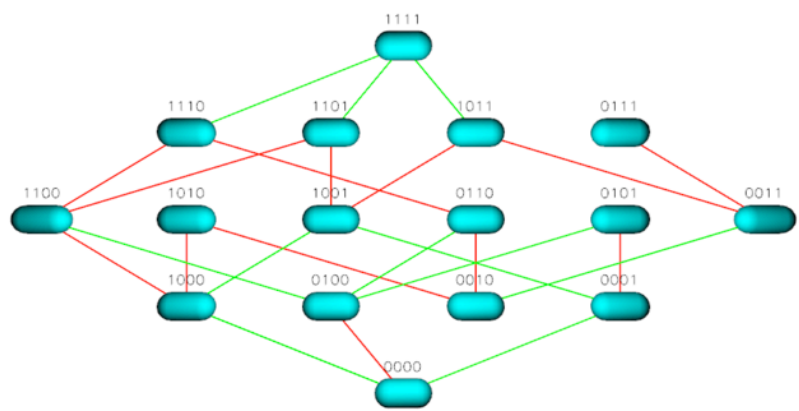

Figure S29 TEM 85 Landscape for Pipercillin + Tazobactam

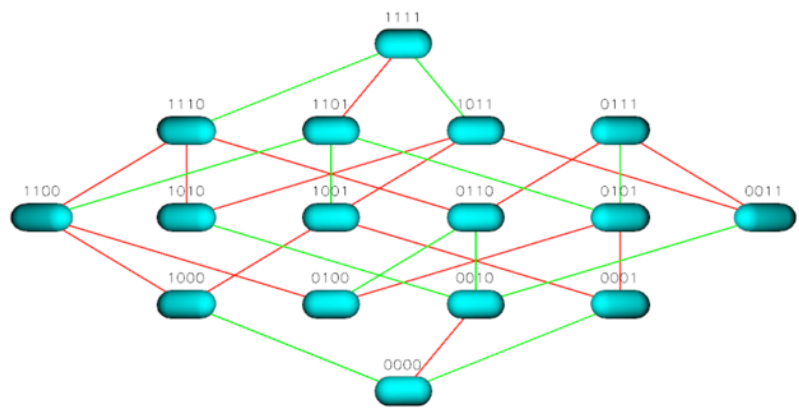

Figure S30 TEM 85 Landscape for Amoxicillin + Clavulanate

Figure S31